# A Comparative Analysis Between SciTokens, Verifiable Credentials, and Smart Contracts: Novel Approaches for Authentication and Secure Access to Scientific Data

MD JOBAIR HOSSAIN FARUK∗, Kennesaw State University, USA*

BILASH SAHA, Kennesaw State University, USA

JIM BASNEY, National Center for Supercomputing Applications, University of Illinois, USA

Abstract: Managing and exchanging sensitive information securely is a paramount concern for the scientific and cybersecurity community. The increasing reliance on computing workflows and digital data transactions requires ensuring that sensitive information is protected from unauthorized access, tampering, or misuse. This research paper presents a comparative analysis of three novel approaches for authenticating and securing access to scientific data: SciTokens, Verifiable Credentials, and Smart Contracts. The aim of this study is to investigate the strengths and weaknesses of each approach from trust, revocation, privacy, and security perspectives. We examine the technical features and privacy and security mechanisms of each technology and provide a comparative synthesis with the proposed model. Through our analysis, we demonstrate that each technology offers unique advantages and limitations, and the integration of these technologies can lead to more secure and efficient solutions for authentication and access to scientific data.

## 1 INTRODUCTION

In the era of digitalization, secure and reliable systems for managing and exchanging sensitive information require increased attention [3]. Every day, an enormous amount of data is generated that is important, particularly in scientific research. Managing credentials such as passwords, secret keys, or stakeholders' sensitive information securely can be a burden to the scientific and cybersecurity communities [5, 8]. With the increasing reliance on computing workflows, digital transactions, and interactions, it has become necessary to ensure that sensitive information is protected from unauthorized access, tampering, or misuse [17]. User authentication and access management is required for various applications such as scientific software systems, network security, Internet of Things (IoT), mobile, web, and cloud-based applications [2]. In general, appropriate security measures should be applied to protect the identity of the stakeholders for trustworthy and digitally verifiable credentials [16]. Applying state-of-the-art technology can prevent the misuse of data.

Several approaches have been introduced to address this need, including SciTokens, Verifiable Credentials, and Smart Contracts, which are all different security technologies for managing and exchanging information securely in a centralized or decentralized environment. SciTokens aims to provide an ecosystem for authorization on distributed scientific computing infrastructures [21]. Verifiable Credentials, a World Wide Web Consortium (W3C) standard, implements the decentralized identity (DID) concept [11], where individuals can prove their identity and qualifications to third parties in a secure and decentralized manner [12]. It enables tamper-resistant claims made by an issuer, where each claim asserts a set of properties or identity attributes about a subject [ 6 ]. Smart Contracts are being adopted in many fields including finance and healthcare [ 18]. Smart contracts refer to a set of programs that are self-verifying, self-executing, and tamper-resistant that integrate with blockchain technology enabling a greater degree of security [ 10, 14]. Smart contracts automate the process of verifying and enforcing the terms of an agreement between parties through centralized data storage and management of credentials and associated keys [1].

Contribution: In this research paper, we aim to provide a comparative analysis of the aforemen- tioned three approaches, (i) SciTokens, (ii) Verifiable Credentials, and (iii) Smart Contracts. We also implement a verifiable credential framework utilizing the smart contract concept, as a case study. We aim to examine the technical features along with the privacy and security mechanisms of each technology and offer a synthesis via the case study. Our analysis provides insights into the applicability of each technology in different contexts and highlights the opportunities and challenges associated with their adoption.

The rest of this paper is organized as follows: Section 2 discusses related work. Section 3 presents our comparative study between the three different approaches. Section 4 introduces our case study of the verifiable credential approach with smart contracts. Section 5 concludes the paper.

## 2 RELATED WORK

In this section, we review the literature and studies related to our research.

SciTokens [21] is a token-based authorization system specifically designed for the scientific community. An OAuth2-based implementation for SciTokens with JSON Web Tokens is supported by scientific software components such as CILogon, CVMFS, HTCondor, and XrootD [22].

M. Sporny et al [20 ] presented the Verifiable Credentials Data Model, which serves as a standard- ized framework for expressing and verifying claims about individuals and entities on the web. The authors discuss the potential use cases for VCs, including identity management and access control in various domains, including scientific data access. In a similar study, Rahma Mukta et al. [16] introduced a novel blockchain-based Self-Sovereign Identity (SSI) platform called CredChain that enables credential sharing. The paper focuses on privacy by utilizing redactable signatures and implementing a selective disclosure scheme. There is also work on a trust model for the Verifiable Credential ecosystem that emphasizes trust between issuers, holders, and verifiers [4].

Related work on smart contract-based authentication includes mobile vehicular networks [23], smart farming [19], smart water environments [13], and IoT [7, 15].

## 3 COMPARATIVE ANALYSIS

Table 1 provides a detailed comparative analysis of SciTokens, Verifiable Credentials, and Smart Contracts. SciTokens provides a token-based authentication and authorization mechanism with JSON Web Tokens (JWT) as the underlying technology. It is compatible with OAuth 2.0 systems and relies on a trusted, centralized token issuer. Access tokens contain user identifiers and permissions, allowing fine-grained control over data access. The token issuer's digital signature ensures the integrity and authenticity of the tokens. However, the centralized nature of the token issuer may create scalability issues and single points of failure in large-scale scientific data systems.

Verifiable Credentials use a decentralized approach to authentication, leveraging cryptographic techniques for attestation and verification [9 ]. They can be stored, managed, and presented by users, offering self-sovereign identity management. Verifiable Credentials utilize digital signatures and can be verified against a registry, such as a distributed ledger (blockchain) or a centralized database. They enable attribute-based authorization, where users can selectively disclose specific attributes when requesting access to scientific data. The performance aspects of Verifiable Credentials are contingent upon the chosen registry solution, and integration complexity may vary accordingly.

Table 1: Comparative analysis between SciTokens, Verifiable Credentials, and Smart Contracts

| Criteria | SciTokens | Verifiable Credentials | Smart Contracts |
| --- | --- | --- | --- |
| Trust | Token issuer is fully trusted | Issuer attestation is provided | Trust established through consensus mechanism |
| Revocation | Short-lived access tokens with revocable refresh tokens | Revocation via the data registry | Revocation logic can be implemented within the contract |
| Privacy | Access tokens contain user identifier (sub) and permissions (scope) | Selective disclosure of attributes | Privacy depends on the contract's design and blockchain features (e.g., zero-knowledge proofs) |
| Security | Tokens digitally signed by issuer | Tokens verified on registry | Security provided by the blockchain |
| Validity | Validity period defined by token issuer | Validity period defined by credential issuer | Validity period and conditions can be set in the contract |
| Verification | Verification relies on the issuer's public key | Verification relies on issuer's attestation and registry | Verification relies on blockchain's consensus and contract logic |
| Authentication | Bearer token authentication | Holder authentication using digital signatures | Contract-based authentication and authorization |
| Functionality | Grants access to named resources/paths | Presents credentials for attribute-based authorization | Grants access to resources based on smart contract terms |
| Scalability | Centralized issuer may become a bottleneck | Scalability depends on the registry implementation | Scalability can be limited by transaction throughput, latency, and contract complexity of the blockchain |
| Interoperability | Interoperable with OAuth 2.0 systems | Interoperable with W3C standards | Interoperable with other contracts and platforms supporting the same standard |
| Ease of integration | Relatively simple integration with existing systems | Integration depends on the chosen registry solution | Integration requires blockchain infrastructure and contract development |
| Credential management | Centralized management by issuer | Management depends on the registry implementation | Decentralized management by the user or contract logic |

Smart Contracts also offer a decentralized approach to authentication and authorization, harnessing the security and consensus mechanisms of blockchain technology. They can be programmed to define access control logic, revocation mechanisms, and conditions under which scientific data can be accessed or shared. Privacy, validity, and authentication features depend on the contract's design, the features of the specific blockchain (e.g., zero-knowledge proofs), and the employed cryptographic techniques. While Smart Contracts provide extensive flexibility and customizability, their integration necessitates more extensive efforts involving blockchain infrastructure and contract development.

## 4 CASE STUDY

For our case study, we developed an application that integrates the Verifiable Credential (VC) and Smart Contract concepts using Ethereum and Hyperledger Fabric to manage credentials containing basic student attributes. Both smart contracts are designed to handle the issuance and verification. Algorithm 1 illustrates the process of the developed Verifiable Credentials with smart contracts.

With Ethereum Smart Contract: The Ethereum implementation uses Solidity to create a custom ERC721 token contract, called StudentCertificate. This contract manages a mapping of certificate IDs to a Certificate struct that contains the certificate's details (name, program, graduation date, GPA, and signature). The issueCertificate function creates a new Certificate struct and computes a hash (using the certificate's details) that serves as the certificate's signature. This function then mints a new ERC721 token corresponding to the certificate. The verifyCertificate function checks the certificate's signature by recomputing the hash using the provided certificate details. If the recomputed hash matches the signature, the certificate is considered verified.

### ALGORITHM 1: Case Study: Issuance and Verification of Verifiable Credentials

procedure Issuance Algorithm (IA)
    Step 1: Input: Name (N), Program (P), Graduation Date (GD), GPA (G)
    Step 2: Concatenate the inputs: $S = N \,||\, P \,||\, GD \,||\, G$
    Step 3: Compute the hash of the concatenated string: $H(S) = h(S)$ where $h$ is a cryptographic hash function (e.g., SHA-256)
    Step 4: Create a new Certificate object (C) containing Name (N), Program (P), Graduation Date (GD), GPA (G), and Signature (Sig), where Sig = H(S)
    Step 5: Store the certificate (C) on the blockchain
end procedure
procedure Verification Algorithm (VA)
    Step 1: Input Certificate (C), Name (N'), Program (P'), Graduation Date (GD'), GPA (G')
    Step 2: Concatenate the inputs: $S' = N' \,||\, P' \,||\, GD' \,||\, G'$
    Step 3: Compute the hash of the concatenated string: $H(S') = h(S')$
    Step 4: Retrieve the stored certificate (C) from the blockchain
    Step 5: Compare the computed hash (H(S')) with the stored hash (Sig) in the certificate (C): $V = (H(S') == Sig)$
end procedure

With Hyperledger Fabric Smart Contract: The Hyperledger Fabric implementation defines an asset, StudentCertificate, containing the certificate's details (name, program, graduation date, GPA, and signature). It also defines two transactions: issueStudentCertificate and verifyStudentCertificate. The issueStudentCertificate transaction function creates a new StudentCertificate asset and computes a hash (using the certificate's details) that serves as the certificate's signature. Finally, the certificate is added to the asset registry. The verifyStudentCertificate transaction function checks the certificate's signature by recomputing the hash using the provided certificate details. If the recomputed hash matches the signature, the certificate is considered verified.

Integration of VC and Smart Contracts: The VC application can interact with both Ethereum and Hyperledger Fabric smart contracts. When the issuer issues a certificate, the application generates a Verifiable Credential payload, including the certificate's details and proof, and encodes it as a JWT token. This JWT token can be shared with any party who wishes to verify the certificate. To verify the certificate, the JWT token is decoded, and the payload is checked against the certificate details. If the payload matches the certificate details, the Verifiable Credential is considered authentic. Addi-tionally, the application calls the verifyCertificate function (Ethereum) or the verifyStudentCertificate transaction (Hyperledger Fabric) to verify the certificate's signature.

From a security and privacy perspective, both implementations in the case study ensure that only valid and authorized entities can issue certificates. The hashing process guarantees data integrity, and the Hyperledger Fabric defines roles and permissions for each participant. This ensures that only authorized participants can issue and verify certificates, thus maintaining the privacy of the data. The Hyperledger Fabric also allows the creation of private communication channels between a specific set of participants, preventing unauthorized access and ensuring privacy. Additionally, transactions must be endorsed by a specified set of peers before being committed to the ledger. This ensures that any changes to the data must be approved by the designated authorities.

Ethereum can include smart contract ownership, which allows only the contract owner to issue certificates. Ethereum uses public key cryptography to secure transactions and ensure the authenticity of the sender. This means that only the holder of the private key can issue transactions, ensuring the integrity of the data and the authenticity of the issuer. Besides, Ethereum uses a consensus mechanism called Proof-of-Work (currently) or Proof-of-Stake (in the upcoming Ethereum 2.0) to secure the network. These consensus mechanisms require participants to prove their commitment to the network, making it difficult for malicious actors to compromise the system.

Employing these security and privacy features to develop Verifiable Credentials with both Hyperledger Fabric and Ethereum ensure that the system remains robust, resistant to tampering, and secure from unauthorized access. However, Hyperledger Fabric appears to be more suitable for the implementation of Verifiable Credentials. The primary reason is that Hyperledger Fabric is a permissioned blockchain, which allows for better access control and privacy management in the context of issuing and verifying credentials. Furthermore, Hyperledger Fabric provides lower transaction costs compared to Ethereum, which is essential for scalability.

## 5 CONCLUSION

In conclusion, the need for secure management and exchange of sensitive information in scientific and cybersecurity communities has led to the development of several approaches such as SciTokens, Verifiable Credentials, and Smart Contracts. In this study, we conducted a comparative analysis of these approaches from various perspectives, including trust, revocation, privacy, and security. We also developed a novel framework for verifiable credentials, leveraging blockchain technology and smart contracts, thereby scrutinizing their technical attributes, privacy, and security mechanisms. Our research findings indicate that while each technology has its distinct advantages and short- comings, integrating smart contracts with verifiable credentials potentially creates more secure and efficient solutions for authenticating and accessing scientific data. Additionally, Hyperledger Fabric delivers robust security and privacy measures. With the incorporation of smart contracts into verifiable credentials, we establish dynamic authentication mechanisms, programmatically enforced, surpassing both SciAuth and JSON Web Tokens in security and efficiency. The decentralized essence of blockchain allows smart contracts to eliminate potential single points of failure, facilitate autonomous multi-factor authentication, and minimize communication overhead, leading to secure, privacy preserved, and seamless access to scientific data.

**ACKNOWLEDGMENTS**

This material is based upon work supported by the National Science Foundation under grant number 2114989.